\documentclass[apjl]{emulateapj}             
\usepackage{graphicx}
\usepackage{txfonts}                        
\usepackage{natbib}
\usepackage{longtable}
\newcommand{\xmm}{XMM--Newton}

\newcommand{\src}{IGR~J16479$-$4514}
\def \ATel {The Astronomer's Telegram}

\submitted{To appear in ApJL, submitted 2008 Apr 14, Accepted May 14}
\shorttitle{{\it Swift} catches an outburst from \src}
\shortauthors{Romano et al.}

\begin{document}
\title{Monitoring Supergiant Fast X-ray Transients with Swift.  \\
Rise to the outburst in IGR~J16479$-$4514. }

\author{P.~Romano\altaffilmark{1}, 
L.~Sidoli\altaffilmark{2}, 
V.~Mangano\altaffilmark{1}, 
S.~Vercellone\altaffilmark{2},  
J.~A.~Kennea\altaffilmark{3},  
G.~Cusumano\altaffilmark{1},  
H.~A.~ Krimm\altaffilmark{4,5}, 
D.~N.~Burrows\altaffilmark{3},  
N.~Gehrels\altaffilmark{6} 
}
  \altaffiltext{1}{INAF, Istituto di Astrofisica Spaziale e Fisica Cosmica, 
	Via U.\ La Malfa 153, I-90146 Palermo, Italy} 
 \altaffiltext{2}{INAF, Istituto di Astrofisica Spaziale e Fisica Cosmica, 
	Via E.\ Bassini 15,   I-20133 Milano,  Italy}
   \altaffiltext{3}{Department of Astronomy and Astrophysics, Pennsylvania State 
             University, University Park, PA 16802, USA}
   \altaffiltext{4}{CRESST/Goddard Space Flight Center, Greenbelt, MD, USA}
 \altaffiltext{5}{Universities Space Research Association, Columbia, MD, USA}
   \altaffiltext{6}{NASA/Goddard Space Flight Center, Greenbelt, MD 20771, USA}


\begin{abstract}

\src\ is a Supergiant Fast X--ray Transient (SFXT), a new class of High
Mass X--ray Binaries, 
whose number is rapidly growing thanks to the INTEGRAL observations 
of the Galactic plane.
It was  regularly monitored with {\it Swift}/XRT since
November 2007, to study the quiescent emission, the outburst properties 
and their recurrence.
A new bright outburst, reaching fluxes above 10$^{-9}$~erg~cm$^{-2}$~s$^{-1}$, 
was caught by the {\it Swift}/BAT. 
{\it Swift} immediately re-pointed at the target with the narrow-field instruments 
so that, for the first time, an outburst from a SFXT
where a periodicity in the outburst recurrence is unknown
could be observed simultaneously in the 0.2--150\,keV energy band. 
The X--ray emission is highly variable and spans almost four orders of
magnitude in count rate during the {\it Swift}/XRT observations covering 
a few days before and after the bright peak. 
The X--ray spectrum in outburst is hard and highly absorbed. 
The power-law fit resulted in a
photon index of 0.98$\pm{0.07}$, and in an absorbing column density of
$\sim$5$\times$10$^{22}$~cm$^{-2}$.
These observations demonstrate that in this source (similarly to
what was observed during the 2007 outburst from the periodic SFXT
IGR~J11215$-$5952), the accretion phase lasts much longer 
than a few hours.
\end{abstract}


\section{Introduction\label{sfxts2:introduction}}

Supergiant Fast X--ray Transients (SFXT) are a new class of High
Mass X--ray Binaries associated with blue supergiant companions, 
several members of which were discovered thanks to the  INTEGRAL observations 
of the Galactic plane  \citep{Sguera2005}.
They are sources with transient X--ray emission  
concentrated in short and bright flares (with a typical duration of a few hours), 
a peak luminosity in the range of 10$^{36}$--10$^{37}$~erg~s$^{-1}$ 
and a quiescent level of 10$^{32}$~erg~s$^{-1}$ \citep[e.g.\ ][]{zand2005}.
The short duration flaring activity is part of a longer accretion phase
at a lower level \citep{Romano2007}.

\src\ is a hard X-ray transient discovered by INTEGRAL in 2003 August 
\citep{Molkov2003:16479-4514}.
Hard X--ray activity was observed on August 8 and 9, at a level of $\sim 12$\,mCrab
(18--25\,keV), while the following day the flux increased by a factor of $\sim$2.
Other outbursts caught with INTEGRAL in 2003 and 
2004--2005 were reported by \citet{Sguera2005} and
 \citet{Sguera2006},  respectively, with peak fluxes above 
10$^{-9}$~erg~cm$^{-2}$~s$^{-1}$ (20--60\,keV). 
The recurrent and short outbursts observed from this source
led to a suggestion that it belongs to the SFXT class.

A frequent hard X--ray (E$>$20 keV) flaring activity  was recently 
discussed by \citet{Walter2007}, who report
on 27 short (duration $<$15~ks) flares and on 
11 long ($>$15~ks) flares in archival INTEGRAL data, spanning times from
2003 January 11 to 2005 December 2, with variable fluxes. 

\src\ was observed once with \xmm\ \citep{Walter2006} in 2004 March, 
when it displayed a low level X--ray emission. 
The joint EPIC pn and ISGRI/INTEGRAL spectrum was 
successfully fit with an absorbed Comptonized spectrum, with
an intrinsic column density of (7.7$\pm{1.7}$)$\times$10$^{22}$~cm$^{-2}$, 
an electron temperature $kT_{\rm e}> 13$\,keV,  and optical depth $\tau <1.8$. 
The unabsorbed 2--100 keV flux was 1.8$\times$10$^{-10}$~erg~cm$^{-2}$~s$^{-1}$. 
An upper limit to the presence of an iron line could be placed
at EW$<280$~eV.

This source is normally detected in the {\it Swift}/BAT transient 
monitor\footnote{http://swift.gsfc.nasa.gov/docs/swift/results/transients/weak/IGRJ16479-4514/} 
(15--50\,keV) 
at a level of 4\,mCrab and showed 23 flares above 300 mCrab and 4.5 $\sigma$  significance during the 
{\it Swift} mission up to 2008 March, including the ones that triggered the {\it Swift}/BAT 
on 2005 August 30 \citep{Kennea2005:16479-4514}, 
2006 May 20 \citep{Markwardt2006:16479-4514}, June 24,  and 2007 July 29. 

\begin{figure}
                \centerline{\includegraphics[width=5cm,height=9cm,angle=270]{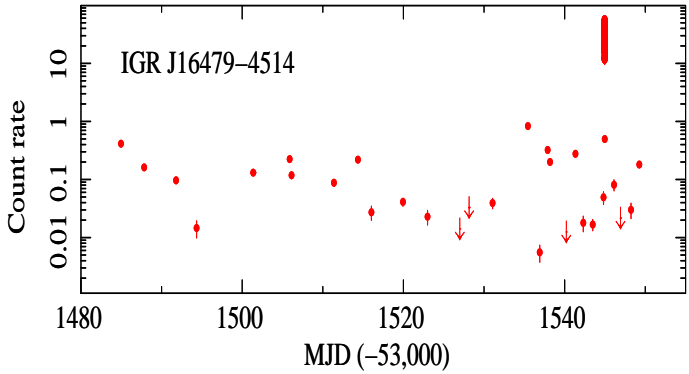}}	
                \caption{{\it Swift}/XRT (0.2--10\,keV) light curve of \src\  in 2008, 
		background-subtracted and corrected for pile-up, PSF losses, and vignetting. 
		Data before 54525 MJD are data reported upon in
		\citet{Sidoli2008:sfxts_paperI}. 
		The downward-pointing arrows are 3-$\sigma$ upper limits. 
 		[See the electronic edition of the
                      Journal for a color version of this figure.]
		                }
                \label{sfxts2:fig:lcv_history}
        \end{figure}

As reported in \citet{Romano2007}, the {\it Swift} monitoring of the outburst of 
the periodic SFXT IGR~J11215$-$5952     in 2007 February represents the deepest 
and most complete set of X--ray observations of an SFXT in outburst. 
We showed that the accretion phase during the bright outburst lasts longer than  
previously thought: days instead of hours, with only the brightest phase
lasting less than one day. 
Stimulated by these {\it Swift} results, we are performing the first sensitive 
X--ray monitoring campaign of the activity of four SFXTs, 
IGR~J16479$-$4514,  IGR~J17391$-$3021, IGR~J17544$-$2619, and IGR~J18410$-$0535,
with the main aim of characterizing their activity on long timescales, both
the quiescent state and the outbursts recurrence, and to 
test our model for the outburst mechanism \citep{Sidoli2007}, 
which involves the presence of a denser and slower equatorial wind component from 
the supergiant companion star. 
The results of the first four months of the on-going  
campaign with {\it Swift}, can be found in a companion paper 
\citep[][Paper I, see Fig.~\ref{sfxts2:fig:lcv_history}]{Sidoli2008:sfxts_paperI}, where 
X--ray activity outside outbursts in \src\ is discussed. 
Here we report on the detailed data analysis of the 
2008 March 19 outburst caught by {\it Swift}/BAT \citep{Barthelmy2008GCN7466} 
and followed at softer energies with {\it Swift}/XRT  
\citep{Romano2008:atel1435}.

\section{Observations and data analysis\label{sfxts2:observations}}

 {\it Swift}/BAT  triggered twice on \src\ on 2008 March 19, 
first at 22:44:47 UT \citep[trigger 306829,][]{Barthelmy2008GCN7466} 
while {\it Swift} had been observing it as part of our monitoring program,
and then at 22:59:59 UT (trigger 306830).  
The spacecraft immediately slewed to the target, so that the narrow-field 
instruments started observing it $\sim 113$\,s after the first trigger
(the slew caused a  $\sim 50$\,s gap in the XRT data, and  the XRT points 
 up to $\sim 60$\,s after the first 
BAT trigger were collected as one of our pointed observations).

The BAT data were analysed using the standard BAT software  
within FTOOLS ({\tt Heasoft}, v.6.4).
Mask-tagged BAT light curves were created in the standard 4 energy bands, 
(Fig.~\ref{sfxts2:fig:lcv_allbands}), 
and rebinned to achieve a signal-to-noise 
of 5. 
BAT mask-weighted spectra were extracted over three time 
intervals strictly simultaneous with XRT data (Sect.~3, Fig.~\ref{sfxts2:fig:lcv_allratios}) 
from the three continuous streams of BAT data in Fig.~\ref{sfxts2:fig:lcv_allbands}. 
Response matrices were generated with {\tt batdrmgen}. 

The XRT data were processed with standard procedures ({\tt xrtpipeline}
v0.11.6), filtering, and screening criteria by using FTOOLS. 
We considered both WT and PC data, 
and selected event grades 0--2 and 0--12, respectively.
To account for the background, we also extracted events within 
source-free regions.
Ancillary response files were generated with {\tt xrtmkarf},
and they account for different extraction regions, vignetting, and
PSF corrections. We used the latest spectral redistribution matrices
(v010) in CALDB. 
All quoted uncertainties are given at 90\% confidence level for one interesting
parameter unless otherwise stated.

\section{Results\label{sfxts2:results}}

\subsection{Light curves}

Figure~\ref{sfxts2:fig:lcv_history} shows the {\it Swift}/XRT 
0.2--10\,keV light curve of \src\ throughout our 2008 monitoring program, 
background-subtracted and corrected for pile-up, PSF losses, 
and vignetting. All data in one segment 
were generally grouped in one point (with the exception of the March 19 
outburst, which shows up as a vertical line on the adopted scale). 
The monitoring program started on 2007 October 26 with approximately 
two observations per week, but when the source showed signs of 
increased activity on March 10, the observations became almost daily. 

\begin{figure}[t]
		\vspace{-1.5truecm}
                \centerline{\includegraphics[width=9.5cm,height=13.5cm,angle=0]{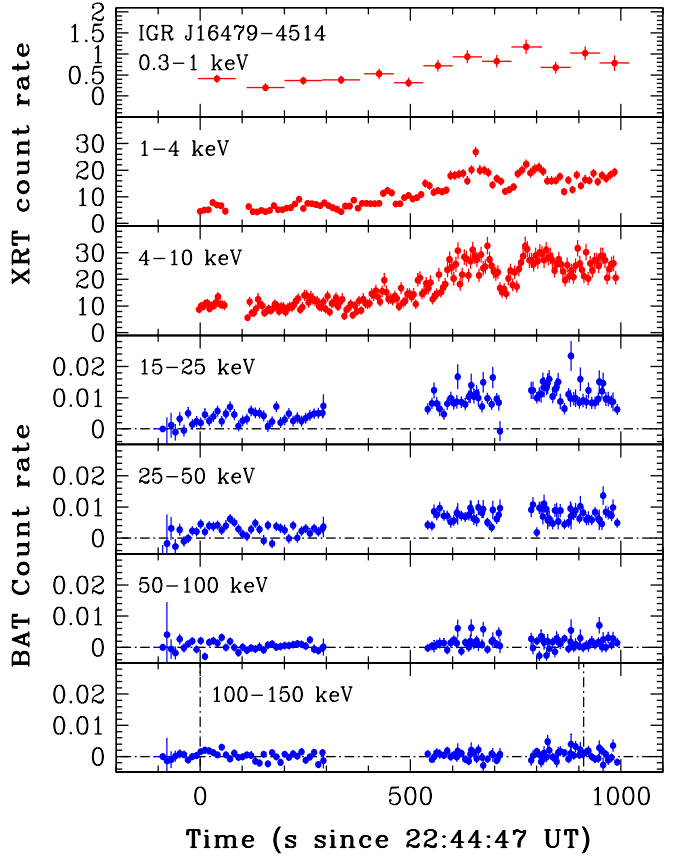}}	
		\vspace{-0.7truecm}
                 \caption{XRT and BAT light curves of the 2008 March 19 outburst 
		in units of count s$^{-1}$ and count s$^{-1}$ detector$^{-1}$, respectively. 
		The XRT points up to $\sim 60$\,s after the first BAT trigger were collected
		 as a pointed observation part of our monitoring program. 
		The gaps in the BAT data are caused by BAT event mode time
		intervals being limited to less than 600\,s to reduce telemetry. 
		The vertical dot-dashed lines in the bottom panel mark the two BAT triggers. 
		 [See the electronic edition of the
                      Journal for a color version of this figure.]
		}
                \label{sfxts2:fig:lcv_allbands}
       \end{figure}

Figure~\ref{sfxts2:fig:lcv_allbands} shows the detailed light curves 
during the brightest part of the March 19 outburst 
in several energy bands. 
%
Figure~\ref{sfxts2:fig:lcv_allratios} shows the 
4--10/0.3--4\,keV, 50--150/15--50\,keV hardness ratios. 
Fitting the 4--10/0.3--4\,keV hardness ratio as a function of time 
to a constant model yields a value of $1.38\pm0.03$ 
and $\chi^2_{\nu}=3.46$ for 8 degrees of freedom (d.o.f.). 
A general trend is observed for a spectral softening as the flux increases 
in the XRT bands.  

During the 2008 March 19 outburst the peak count rate exceeded 
the one recorded on 2005 August 30 
\citep{Kennea2005:16479-4514,Sidoli2008:sfxts_paperI} by a factor of $\sim 5$. 
The total (0.3--10\,keV)  XRT light curve 
(Fig.~\ref{sfxts2:fig:lcv_allratios}a)
shows an increase in count rate by a factor of 
$\sim 5$ in  $\sim 13$ minutes, and an increase by a factor of $\sim 350$ in  $\sim 3$ hours
(comparing with the first earliest point, not shown in Fig.~\ref{sfxts2:fig:lcv_allratios}). 
A corresponding increase is observed in the BAT flux on timescales of minutes.

A timing analysis was performed on WT data ($\Delta T=884$\,s), 
after having converted the event arrival times to the Solar System 
Barycentric frame. We searched for coherent periodicity, 
but found no evidence in the range 10\,ms--100\,s.

\subsection{Spectra} 

We extracted the mean spectrum of the brightest X-ray emission 
(obs.\ 00306829000) 
and performed a fit in the 0.3--10\,keV band of the WT data, 
which were rebinned with a minimum of 20 counts bin$^{-1}$ to allow $\chi^2$ fitting. 
An absorbed power-law model yielded an absorbing column of 
$N_{\rm H}=(5.51_{-0.28}^{+0.29})\times 10^{22}$ cm$^{-2}$, 
a photon index $\Gamma=0.98\pm0.07$, and 
$\chi^2_{\nu}=0.939$ (563 d.o.f., see Table~\ref{sfxts2:tab:specfits}).
The unabsorbed flux in the 2--10\,keV band is $5.8\times10^{-9}$ erg cm$^{-2}$ s$^{-1}$. 
A high-energy cutoff power-law model ({\tt cutoffpl} in XSPEC) 
yielded  $N_{\rm H}=(4.65_{-0.53}^{+0.55})\times 10^{22}$ cm$^{-2}$, 
$\Gamma=0.16\pm0.48$, $E_{\rm c}=7_{-3}^{+10}$, $\chi^2_{\nu}=0.927$ (562 d.o.f.). 
The Ftest probability with respect to the absorbed power-law model is 
$3.154\times10^{-3}$ (2.95 $\sigma$), hence, since the cutoff energy is 
not well constrained, we favour the absorbed power-law model for the XRT data alone.
We note however that a cutoff power-law yields a good fit to the joint XRT+BAT data.
We also note that the derived $N_{\rm H}$ is in excess of the one along the line of sight,
$1.87\times 10^{22}$ cm$^{-2}$.

The X-ray spectrum of the fainter emission (PC data, unabsorbed  2--10\,keV flux 
of $8.95\times10^{-11}$ erg cm$^{-2}$ s$^{-1}$) is significantly softer: 
fitted using Cash statistics and adopting an absorbed power-law model model, 
we obtain $\Gamma=2.57_{-0.68}^{+0.75}$ and  
$N_{\rm H}=(5.76_{-1.71}^{+2.05})\times 10^{22}$ cm$^{-2}$ (C-stat$=471.0$ for 68.93\% of 10$^4$ 
Monte Carlo realizations with statistics $<$ C-stat).  

\begin{figure}[t]
                \centerline{\includegraphics[width=6cm,height=10cm,angle=270]{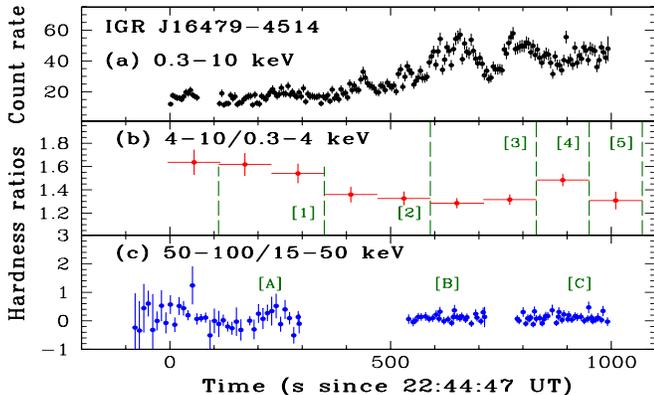}}	
			\vspace{-0.5truecm}
                \caption{X-ray WT mode and BAT event mode hardness ratios 
		during the 2008 March 19 outburst. 
		{\bf (a):}   XRT 0.3--10\,keV light curve. 
		{\bf (b):}  Ratio of the 4--10 and 0.3--4\,keV XRT light curves.
		The vertical dashed lines mark the five intervals for XRT time-resolved spectroscopy
		(Table~\ref{sfxts2:tab:specfits}). 
		{\bf (c):} Ratio of the 50--150 and 15--50\,keV BAT light curves. 
		[A], [B], and [C] mark the three intervals for joint XRT+BAT spectroscopy (Table~\ref{sfxts2:tab:specfits2}).
 		[See the electronic edition of the
                      Journal for a color version of this figure.]
		   }
                \label{sfxts2:fig:lcv_allratios}
        \end{figure}

Upon examination of the best XRT hardness ratio (Fig.~\ref{sfxts2:fig:lcv_allratios}b),  
we accumulated spectra in five time bins which maximized the variations in 
hardness ratio and yielded $\sim 2600$--7500 counts per spectrum. 
Table~\ref{sfxts2:tab:specfits} reports our results.

\begin{figure}[t]
                \centerline{\includegraphics[width=5cm,height=9cm,angle=270]{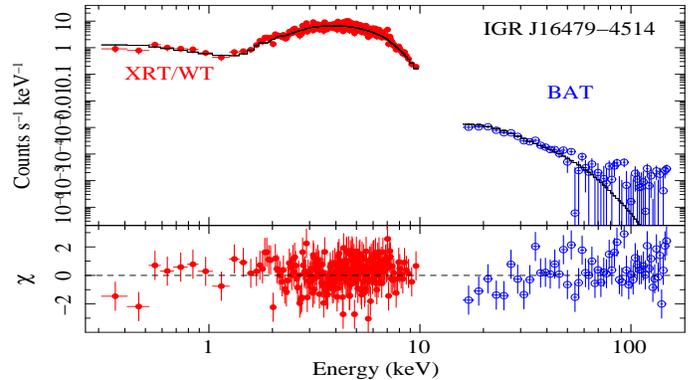}}	
                \caption{Spectroscopy of the 2008 March 19 outburst. 
		{\bf Top:} data from the second BAT observation 
		(part [C] in Fig.~\ref{sfxts2:fig:lcv_allratios}) 
		and simultaneous XRT/WT data
		fit with an absorbed power law with a high energy cutoff. 
		{\bf Bottom:} the residuals of the fit (in units of standard deviations). 
 		[See the electronic edition of the
                      Journal for a color version of this figure.]
		                }
                \label{sfxts2:fig:meanspec}
        \end{figure}

We extracted three XRT spectra simultaneous with the three BAT spectra 
in the time intervals  120--303\,s,  541--723\,s, and  793--1092\,s since the BAT
trigger ([A], [B], and [C] in Fig.~\ref{sfxts2:fig:lcv_allratios}), 
and performed joint fits in the 
0.3--10\,keV and 14--150\,keV energy bands, for XRT and BAT, respectively,
and applied an energy-dependent systematic error vector to the BAT data. 
Factors were included in the fitting to allow for normalization
uncertainties between the two instruments, which were constrained to be within their 
usual ranges (0.9--1.1). 
Several models typically used to describe the X--ray emission from 
accreting pulsars in HMXBs were adopted. 
The results are reported in Table~\ref{sfxts2:tab:specfits2} and 
Fig.~\ref{sfxts2:fig:meanspec}. 
We note that the fit to the BAT data alone obviously results in a steep power law
with $\Gamma\sim 3$. 

All models allow a good deconvolution of the 0.5--100 keV emission,
resulting in a rather flat continuum below 10\,keV together with a
high--energy exponential cutoff (model {\tt highecut} in XSPEC) 
at $\sim$8--20\,keV (depending on the
time interval). 
In the Comptonization model ({\tt compTT} in XSPEC, 
\citealt{Titarchuk1994}), we adopted a spherical geometry for the
Comptonizing plasma, since both geometries allow an equally good
deconvolution.

\section{Discussion\label{sfxts2:discussion}}

\begin{deluxetable}{llrrrr}
  \tabletypesize{\scriptsize}
  \tablewidth{0pc} 	      	
  \tablecaption{Spectral fits of XRT data.\label{sfxts2:tab:specfits}} 
  \tablehead{
\colhead{Spectrum} & \colhead{Mid Time} & \colhead{$N_{\rm H}$} & \colhead{$\Gamma$} &  \colhead{$\chi^2_{\nu}$ (d.o.f.)/} \\
\colhead{}       & \colhead{(MJD) } & \colhead{(10$^{22}$ cm$^{-2}$)}    & \colhead{} & \colhead{ C-stat (\%)} 
}
  \startdata
Total WT  & 54544.9551 &  $5.51_{-0.28}^{+0.29}$   &  $0.98_{-0.07}^{+0.07}$    &  $0.939$  $(563)$  \\
Part 1 & 54544.9511 &  $6.05_{-1.00}^{+1.19}$	&  $0.83_{-0.24}^{+0.26}$    &  $0.947$  $(95)$    \\
Part 2 & 54544.9522 &  $4.92_{-0.62}^{+0.70}$	&  $0.77_{-0.16}^{+0.17}$    &  $0.997$  $(149)$   \\
Part 3 & 54544.9555 &  $5.07_{-0.41}^{+0.45}$	&  $1.02_{-0.12}^{+0.12}$    &  $0.982$  $(249)$   \\
Part 4 & 54544.9569 &  $5.50_{-0.62}^{+0.69}$	&  $0.93_{-0.16}^{+0.17}$    &  $0.820$  $(157)$   \\
Part 5 & 54544.9596 &  $5.91_{-0.71}^{+0.82}$	&  $1.12_{-0.18}^{+0.19}$    &  $0.796$  $(128)$   \\
Total PC  & 54544.9986 &  $5.76_{-1.71}^{+2.05}$   &  $2.57_{-0.68}^{+0.75}$    &  471.0  $(68.93)$\tablenotemark{a} 
  \enddata 
  \tablenotetext{a}{Cash statistics and percentage of Monte Carlo realizations with statistic $<$C-stat. 
}
  \end{deluxetable}

We report on the first simultaneous broad band X--ray spectrum of \src\ 
in the 0.3--100\,keV energy range. 
The source emission is well fit with the spectral models
usually applied to the accreting X--ray pulsars: power laws with a high 
energy cutoff \citep{White1983} and Comptonization models 
\citep{Titarchuk1994}.
The resulting parameters are also very similar to those of this
kind of X--ray binary sources. 

The time-resolved spectroscopy of the joint XRT--BAT spectrum during the
bright flare reveals that the initial part of the flare (part [A]) shows
the flattest continuum below 10\,keV, together with the lowest
high energy cutoff, with respect to the subsequent brighter part of the
emission (see Table~\ref{sfxts2:tab:specfits2}, parts [B] and [C]), 
in all the three models adopted.
No evidence is found for variability in  the absorbing column density
during the evolution of the bright flare,
and also in comparison with the out-of-outburst emission \citep{Sidoli2008:sfxts_paperI}.
From optical/IR observations of the supergiant companion, Rahoui et al.
(2008) derived
an extinction A$_{\rm V}=18.5$\,mag, which translates into
an absorbing column density of $3.3\times$10$^{22}$~cm$^{-2}$
\citep{avnh} which is in excess of the total Galactic in the source
direction (see Sect. 3). 
Thus there is evidence for neutral absorbing matter local to the binary system. 
Among the different models used to fit the broad-band spectrum,
the Comptonization model yieded an $N_{\rm H}$ compatible 
with the optical extinction, 
thus there is no strong evidence that an excess of absorption
is local to the X--ray source (and that could be linked to the 
accreting clumps, if the supergiant wind is inhomogeneous).
This could imply that the bulk of the neutral absorbing matter is
circum-binary matter, instead of being associated with the 
accreting clumps (whose matter is very likely totally ionized when 
approaching the X--ray source).
Although the statistics allow a constant value of  $N_{\rm H}$ 
(Table~\ref{sfxts2:tab:specfits2}), a trend is found 
in the hardness ratio (Fig.~\ref{sfxts2:fig:lcv_allratios}b), 
which is decreasing with increasing flux, during the
outburst. This is probably the indication of a reduced neutral gas column density,
as the circumstellar gas is progressively ionized by the X--ray flare.

Before these observations, IGR~J11215$-$5952
was the only SFXT observed in depth during an outburst 
(\citealt{Romano2007}, \citealt{Sidoli2007}).
In that case, the occurrence of the outburst was predictable thanks
to the periodic recurrence \citep{SidoliPM2006}.
For \src\ the outburst could only be caught thanks 
to our {\it Swift} monitoring campaign 
\citep{Sidoli2008:sfxts_paperI}.

The source light curve 
before 54525 MJD (Fig.~\ref{sfxts2:fig:lcv_history}) 
has already been reported and discussed in \citet{Sidoli2008:sfxts_paperI}. 
It shows a smoothly variable flux, with
a dynamic range of more than one order of magnitude, which apparently is
sinusoidally modulated with a period of $\sim$24~days (although we
caution that only about 2 periods of this suggested periodicity have been
covered).
During the most frequent monitoring (between 54535 and 54544~MJD)
preceding the outburst, a highly variable source intensity was observed, 
with rates going up and down on short time scales of 1--2 days and 
spanning two orders of magnitude at maximum, before reaching the
brightest peak.
In the total light curve monitored with {\it Swift}, \src\ spanned
almost four orders of magnitudes in flux in a few days.
Moreover, the flux is highly variable on different time scales, from
seconds to minutes, days and weeks,
revealing a very intense flaring activity, both during the outburst and
outside it \citep{Sidoli2008:sfxts_paperI}. 

The rise time to the peak (at an average rate of $\sim$17\,counts~s$^{-1}$ on
54544.95 MJD)
is at most $\sim$3~hrs. After the peak (also monitored by the BAT),
the nearest XRT observation ($\sim 1.175$\,days later) found 
the source at a much fainter rate (0.085$\pm0.017$ counts s$^{-1}$). 
This implies that the brightest phase of the outburst lasts one day at
most, which is a behaviour similar to what observed in IGR~J11215$-$5952 
\citep{Romano2007}, where the bright emission lasted less than 1 day, 
and then was followed by a much weaker emission, although highly variable, 
with several shorter (but fainter) flares. 
This is also consistent with the duration of previous outbursts detected by the BAT.

\begin{deluxetable}{lrrrrr}
  \tabletypesize{\scriptsize}
  \tablewidth{0pc} 	      	
  \tablecaption{Spectral fits of simultaneous XRT and BAT data.\label{sfxts2:tab:specfits2}}
  \tablehead{
\colhead{Spectrum} & \colhead{} & \colhead{} & \colhead{Parameters} &  \colhead{} &  \colhead{} 
}
  \startdata
{\sc highecutpl}\tablenotemark{a} &  $N_{\rm H}$  & $\Gamma$  &$E_{\rm c}$ (keV) &$E_{\rm f}$ (keV) & $\chi^{2}_{\nu}$ (dof)  \\
\noalign{\smallskip\hrule\smallskip}
Part A 		& 5.6$^{+1.2}_{-0.96}$	&0.74$^{+0.29}_{-0.29}$  &7.2$^{+2.2}_{-1.5}$ &9.9$^{+2.9}_{-1.6}$  & 0.823 (150)   \\
Part B 		& 5.7$^{+0.4}_{-0.6}$	&1.24$^{+0.11}_{-0.17}$  &6.9$\pm{1.0}$       &17.7$^{+2.8}_{-3.2}$ & 0.982 (264)  \\ 
Part C 		& 6.2$^{+0.6}_{-0.5}$ 	&1.15$^{+0.15}_{-0.13}$  &6.6$^{+1.0}_{-0.7}$ &15.3$^{+2.7}_{-1.8}$ &1.046 (301) \\

\noalign{\smallskip\hrule\smallskip}
                 & $L_{\rm 0.5-10}$\tablenotemark{c}    &$L_{\rm 0.5-100}$\tablenotemark{c}   & &  \\
\noalign{\smallskip\hrule\smallskip}
Part A 		& 0.94 & 2.05 & \nodata& \nodata & \nodata \\ 
Part B 		& 2.66 & 5.46 & \nodata& \nodata & \nodata \\ 
Part C 		& 2.82 & 5.73 & \nodata& \nodata & \nodata \\ 
\noalign{\smallskip\hrule\smallskip}
{\sc cutoffpl}\tablenotemark{a}   & $N_{\rm H}$  & $\Gamma$  & $E_{\rm c}$ (keV) &  & $\chi^{2}_{\nu}$ (dof)  \\
\noalign{\smallskip\hrule\smallskip}
Part A 		& 5.4$^{+1.2}_{-1.1}$ &0.17$\pm$0.40   &7.4$^{+2.3}_{-1.5}$  & \nodata &0.845 (151)  \\ 
Part B 		& 5.8$^{+0.6}_{-0.5}$ &1.0$\pm{0.2}$   &15.3$^{+3.5}_{-2.6}$ & \nodata &1.042 (265)   \\
Part C 		& 6.5$\pm{0.6}$       &0.97$\pm{0.17}$ &13.5$^{+2.5}_{-1.9}$ & \nodata &1.108 (302)  \\
\noalign{\smallskip\hrule\smallskip}

{\sc compTT}\tablenotemark{b} & $N_{\rm H}$   & $T_{\rm 0}$ (keV)      &$T_{\rm e}$ (keV)    & $\tau$      & $\chi^{2}_{\nu}$ (dof) \\
\noalign{\smallskip\hrule\smallskip}
Part A 		&3.2$^{+1.0}_{-0.7}$ & 1.57$^{+0.24}_{-0.30}$ &7.1$^{+3.3}_{-2.0}$  & 7.8$^{+4.2}_{-3.8}$ &0.829 (150)  \\
Part B 		&2.8$\pm{0.4}$       & 1.36$\pm$0.11          &12.0$^{+4.0}_{-2.2}$ & 5.3$^{+1.0}_{-1.4}$ & 0.962 (264)  \\
Part C 		&3.2$\pm{0.4}$       & 1.41$^{+0.12}_{-0.10}$ &10.8$^{+6.4}_{-1.5}$ & 5.5$^{+0.3}_{-2.2}$ &1.00 (301) 
\smallskip
  \enddata 
  \tablenotetext{a}{$N_{\rm H}$ is the neutral hydrogen column density ($\times 10^{22}$ cm$^{-2}$), 
                          $\Gamma$ the power law photon index, $E_{\rm c}$ the cutoff energy (keV), $E_{\rm f}$
                         the exponential folding energy (keV).}
  \tablenotetext{b}{$T_{\rm 0}$ is the temperature of the Comptonized seed photons, $T_{\rm e}$ and $\tau$ are  
		the temperature and the optical depth of the Comptonizing electron plasma (spherical geometry).}
  \tablenotetext{c}{In units of 10$^{37}$ erg s$^{-1}$ derived  assuming a distance of 4.9~kpc. 
			}
  \end{deluxetable}

\citet{Rahoui2008} classify the optical companion as an O8.5I star
located at $\sim$4.9~kpc.
This, together with the highly variable X--ray flux, confirms the
classification as a SFXT. 
In these sources, a subclass of HMXBs, 
the accretion onto the compact object is very likely through the strong
wind from the supergiant donor. 
The orbital parameters of \src\ as well as the nature of the compact
object are still unknown,
although the X--ray spectral properties are similar to those of accreting
pulsars.
Thus, the compact object is probably a neutron star, as in
IGR~J11215$-$5952,
which hosts a pulsar with a period of $\sim$187~s
\citep{Swank2007:atel999}.

Assuming the compact object is a neutron star and 
the typical parameters for an O-type supergiant (mass, $M\sim50$--$60~M_\odot$,
a radius $R=25$--$30~R_\odot$, a beta-law for the supergiant wind with
an exponent $\beta=1$,
a wind terminal velocity of 1800--2200\,km\,s$^{-1}$, and a
wind mass loss of 2.5$\times$10$^{-6}$\,$M_\odot$~yrs$^{-1}$; 
\citealt{Wilson1985}), we can compare the expected
X--ray luminosity for Bondi-Hoyle accretion with the observed X--ray
emission.
The fast  ($<3$\,hrs)  rise to the brightest emission during outburst
is difficult to explain with 
an enhanced accretion rate when the compact object approaches the O-type
supergiant
companion, along its orbit, if we assume a symmetric and homogeneous wind.
The same behaviour was observed in the periodic
IGR~J11215$-$5952  \citep{Sidoli2007}.
In this framework, the fainter X--ray emission
(10$^{34}$--10$^{35}$~erg~$^{-1}$)
preceding the bright outburst cannot be explained with an orbit with a
period shorter
than $\sim15$\,days, and an eccentricity lower than $\sim$0.5.
A more detailed light curve modeling cannot be performed 
at this time, since the orbital parameters are currently unknown.

{\it Facilities:} \facility{{\it Swift}}.

\acknowledgements
We thank the {\it Swift} Team for making these observations possible,
particularly the duty scientists and science planners.
PR thanks INAF-IASF Milano, where part of the work was carried out, for their kind hospitality. 
This work was supported in Italy by MIUR grant 2005-025417 and contract ASI/INAF I/023/05/0, I/088/06/0,  
and at PSU by NASA contract NAS5-00136.


\end{document}